\def\year{2022}\relax
\documentclass[letterpaper]{article} % DO NOT CHANGE THIS
\usepackage{aaai22}  % DO NOT CHANGE THIS
\usepackage{times}  % DO NOT CHANGE THIS
\usepackage{helvet}  % DO NOT CHANGE THIS
\usepackage{courier}  % DO NOT CHANGE THIS
\usepackage[hyphens]{url}  % DO NOT CHANGE THIS
\usepackage{graphicx} % DO NOT CHANGE THIS
\usepackage{natbib}  % DO NOT CHANGE THIS AND DO NOT ADD ANY OPTIONS TO IT
\usepackage{caption} % DO NOT CHANGE THIS AND DO NOT ADD ANY OPTIONS TO IT
\DeclareCaptionStyle{ruled}{labelfont=normalfont,labelsep=colon,strut=off} % DO NOT CHANGE THIS
\usepackage{algorithm}
\usepackage{algorithmic}
\usepackage{mathtools}
\mathtoolsset{showonlyrefs}

\usepackage{newfloat}
\usepackage{listings}
\floatstyle{ruled}
\newfloat{listing}{tb}{lst}{}
\floatname{listing}{Listing}
\usepackage{dsfont}
\usepackage{amsfonts}
\usepackage{amsmath}

\title{Influence of Repetition through Limited Recall}
\author {
    Jad Sassine\textsuperscript{\rm 1},
    M. Amin Rahimian\textsuperscript{\rm 2},
    Dean Eckles\textsuperscript{\rm 1}
}
\affiliations {
    \textsuperscript{\rm 1} Massachusetts Institute of Technology\\
    \textsuperscript{\rm 2} University of Pittsburgh\\
    jsassine@mit.edu, rahimian@pitt.edu, eckles@mit.edu\\
    \vspace{-50pt}
}

\usepackage{bibentry}
\begin{document}

\maketitle

\begin{abstract}
Decision makers who receive many signals are subject to imperfect recall. This is especially important when learning from feeds that aggregate messages from many senders on social media platforms. In this paper, we study a stylized model of learning from feeds and highlight the inefficiencies that arise due to imperfect recall. In our model, failure to recall a specific message comes from the accumulation of messages which creates interference. We characterize the influence of each sender according to the rate at which she sends messages and to the strength of interference. Our analysis indicates that imperfect recall not only leads to double-counting and extreme opinions in finite populations, but also impedes the ability of the receiver to learn the true state as the population of the senders increases. We estimate the strength of interference in an online experiment where participants are exposed to (non-informative) repeated messages and they need to estimate the opinion of others. Results show that interference plays a significant role and is weaker among participants who disagree with each other. Our work has implication for the diffusion of information in networks, especially when it is false because it is shared and repeated more than true information.
\end{abstract}

\section{Introduction}

People increasingly get information via  \emph{feeds} that aggregate messages from many senders into a list. The role of these feeds in opinion formation and decision making has been the subject of widespread speculation and some empirical research. For example, commentators argue that these feeds create or facilitate “echo chambers” or “filter bubbles” \citep{pariser2011filter}. While empirical evidence gives a more mixed picture — in which social media exposes individuals to both information sources that share or do not share their ideological and partisan alignment \citep{bakshy2015exposure, flaxman2016filter, eady2019many}, these concerns highlight the importance of the combination of sender behavior, platform policies, and receiver information processing in the evolution of public opinion. 

In particular, senders are often trying to change the opinions of others, which can motivate them to repeatedly share related messages (e.g., by fake news sites, politicians, and advertisers). While repetition perhaps should not be a problem in theory, in practice individuals are subject to double counting: they do not consider the correlation structure of the signals that they receive. This process can lead to a reinforcing feedback loop where repeated information leads to extreme opinion, which leads to further repeated information and more extreme opinion. How should platforms modulate exposure to repeated messages in order to facilitate more accurate information processing? Answering this question benefits from positing and validating a precise mechanism behind double counting. 

There are two popular explanations behind double-counting. One explanation is that the task is conceptually complex: individuals fail to realize that taking the correlated signals at their face value is a problem or that they can back out the independent signals. The second explanation is that the task is mathematically complex, i.e., people know what they need to solve, but they do not know how to solve it. In an experiment where participant need to estimate an ex ante unknown state of the world from correlated signals, \citet{enke2017correlation} find that participants struggle primarily with noticing the presence of correlation and understanding how this can be dealt with, both of which can be viewed as conceptual problems. However, noticing the problem and understanding how to solve it are two distinct mechanisms and, based on this experiment, the importance of one relative to the other is unclear. In this paper, we investigate individuals’ failure to notice repetition through the lens of limited memory: people double count because they fail to keep track of the source of messages that they have received. In other words, they have imperfect recall of the sources of the signals that have shaped their beliefs.

To study the effect of limited recall on learning, we propose a simple model that is both analytically tractable and accounts for two empirically validated facts of human memory \citep{kahana2012foundations}. Repetition increases the likelihood of recall. However, it is constrained by what we call interference: messages from other senders decrease the probability of remembering any single message. This model naturally lends itself to the study of variations in the rate of activity and number of different sources. 

The purpose of this paper is to show the implications of limited recall on learning, especially when variation in user activity affects the recall process. We analyze a stylized model of feeds. The unknown state of the world takes one of two values: zero or one. The environment consists of $n$ sources, each of whom is endowed with a private signal that is a Bernoulli random variable. We assume that private signals are independent conditioned on the true state, and they are informative about the true state (e.g., each signal is more likely to agree with the true state of the world than not). Each source transmits her private signal to the user (henceforth, the “receiver”), repeatedly, at random times according to a fixed-rate, Poisson point process. The user feed consists of the list of such messages, ordered in time. We study the quality of information aggregation as the receiver is repeatedly exposed to the private signals of different sources. The non-Bayesian belief (subject to interference and imperfect recall) exhibits certain inefficiencies: (i) It concentrates on one of the two states, despite limited information from finite $n$ sources as time advances to infinity; i.e., the non-Bayesian receiver holds more extreme opinions, the more she interacts with her feed. (ii) The limiting belief may concentrate on a false state and is disproportionately affected by different users according to their transmission rates. Our analytical formulation allows us to quantify the influence of each sender, as a function of a parameter that can be interpreted as the strength of interference. 

To estimate this parameter, we run an online experiments where we show respondents real comments that were taken from the web about a picture ``The Dress'' that generated significant disagreement when it first came out in February 2015 \citep{lafer2015striking}. In fact, different people see different colors: some people see the dress as black and blue; others see it as white and gold. After showing a set of messages, we ask participants to answer `how many people believe the color is ...' for both `black and blue' and `white and gold'? and pay them according to their performance. In our setup, participants see three different feeds. In the first feed, they do not know what the question is. After they answer the question at the end of the first feed, participants are shown an additional two feeds. The randomization occurs for each feed, leading to within (in addition to between) subject randomization. This design choice serves both as training and allows us to compare how people learn when they are passive (as opposed to active in the second and third feed). The second experimental variation is repetition: some messages are repeated from one to six times, allowing to see the effect of increasing repetition on errors in both questions.

This experimental setup has several advantages. First, the `objective' nature of the perceptual differences implies that repeated messages contain no additional informational content. In fact, one of the biggest issues with over-counting is that repetition may be interpreted by the respondent as evidence that the sender has learned something new about the world or his opinion has changed. For example, one's opinion of a painting may change over repeated exposures. Here, the picture is such that the color people see remains fixed throughout. As a result, seeing the message `I see the color white and gold' twice cannot be rationally interpreted as more than a single piece of evidence. Second, the goal of the agent is to count the distinct number of senders, not update her beliefs as is standard in most studies of probabilistic reasoning and judgment biases; for a review see \citet{benjamin2019errors}. This allows us to abstract away from the inability of individuals to perform the Bayesian update, making the strength of interference more identifiable.

Results show that overcounting as a function of repetition increases at a decreasing rate, reaching a maximum average of 1 message overcounted for six repetitions. This overcounting is lower when participants disagree with the high sender or when they know specifically the question they will be asked (feeds two and three). Our work has implications for settings in which people form opinions or make decisions based on the opinion of others \citep[e.g.,][]{moehring2021surfacing, stewart2019information}. Our results show that even though people are exposed to ideologically diverse opinions \citep{bakshy2015exposure} this does not prevent one side from exerting more influence through repetition. This issue is particularly notable in light of new evidence that fake news are shared more than their true counterparts \citep{vosoughi2018spread}, and such sharing is concentrated among a small portion of the population \citep[e.g.,][]{grinberg2019fake}. This suggests policies aimed at not just exposing users to diverse information but also specifically limiting messages from users who disproportionately repeat themselves. 
\vspace{-10pt}

\section{Model}\label{sec:model}
 We consider a sequential learning model, where the agent (referred to as the receiver) receives a message at random times according to a Poisson point process. A message at a random time $\mathbf{t}$ is a random vector comprised of source and content: $\mathbf{m}_{\mathbf{t}} = (\mathbf{o}_{\mathbf{t}},\mathbf{c}_{\mathbf{t}})$. Both $\mathbf{m}_{\mathbf{t}}^{(1)} := \mathbf{o}_{\mathbf{t}}$ and $\mathbf{m}_{\mathbf{t}}^{(2)} := \mathbf{c}_{\mathbf{t}}$ are random variables. Upon reception of a message, the receiver updates her belief unless she has received the same message (i.e., the same content from the same source) in the past and she remembers it. Note that whereas \citet{enke2017correlation} consider correlated but different signals, in this paper we are only concerned with the repetition of the same signal. This more stylized setup can be interpreted as extreme correlation. 
 
 The sequence of messages constitute the receiver's feed. Here we study how imperfect recall affects the receiver's ability to learn from her feed.

\subsection{Modeling Memory}

Our model draws on the literature of cued recall which suggests that, upon the reception of a cue, previously received messages are recalled proportionally to similarity with the cue \citep{kahana2012foundations,bordalo2017memory,NBERw29273}. This implies that (1) repetition leads to a higher probability of recall, since there are more messages and therefore more chances to recall the sender of the message, and (2) other messages (e.g., closer in time) create interference: because of their similarity with the cue (closeness in time), they are more likely to be recalled and therefore reduce the probability of recalling any given sender \citep{NBERw29273}. Based on these two ideas, we model the probability that a source $i$ is recalled at time $t$ as:
\begin{multline}
    \mathbb{P}(\mbox{$i$ recalled at $t$}) = \\ \frac{|\tau<t, \mathbf{m}_{\tau}^{(1)} = i|}{|\tau<t, \mathbf{m}_{\tau}^{(1)} = i| + r|\tau<t, \mathbf{m}_{\tau}^{(1)} \neq i|},
    \label{eq:proba_remember}
\end{multline} where $|\tau<t, \mathbf{m}_{\tau}^{(1)} = i|$ corresponds to the number of messages sent by $i$ before $t$, $|\tau<t, \mathbf{m}_{\tau}^{(1)} \neq i|$ to the total number of messages sent by other senders before $t$ and $r \geq 0$ is a parameter corresponding to the strength of interference. Note that $r$ can also be interpreted as the degree to which other messages are dissimilar to the message sent by $i$. If there is no interference, then $r = 0$. 

This formulation has desirable properties. If $m$ corresponds to the total number of messages sent by $i$ and $T$ to the total number of messages sent by other senders, then the probability that source $i$ is recalled increases at rate $rT/(m+rT)^2$, which decreases with $m$. As $T \to \infty$ so that there are many messages from other senders and the probability of recall is very low, this rate is approximately constant in $m$ and the sender expects the benefits of their repetition to diminish as the probability of recall increases linearly with $m$. On the other hand, if $T \to 0$ or $r \to 0$, then the probability of remembering $i$ is high but not sensitive to $m$, so the sender can expect a small, fixed benefit in this regime.  

\section{Analytical Results in the Binary Environment}\label{sec:analysis}

Consider a continuous time $t>0$ and a binary state of the world, denoted by $\theta \in \{0,1\}$ that is unknown to the receiver. We have $n$ sources of information, labeled by $i \in \{1,\ldots,n\}$. Each source is endowed with a private signal that is an independent Bernoulli random variable, denoted by $\mathbf{s}_i$ whose distribution is given by: $\mathbb{P}(\mathbf{s}_i = \theta) = \overline{p}_i > \mathbb{P}(\mathbf{s}_i = 1 - \theta) = \underline{p}_i$.

Each source ($i$) transmits with a fixed rate $\alpha_i$ at random times according to a Poisson point process. Each transmission by the source $i$ at a time $\mathbf{t} \in \mathbb{R}^{>0}$ constitutes a message of the form $\mathbf{m}_{\mathbf{t}} = (i,\mathbf{s}_{i})$ that appears in the user's feed.  We also use the notation $\bar{\alpha}_n = \sum_{i=1}^{n}\alpha_i$ for the sum of the rates. 

We study the learning outcome by analyzing the asymptotic  belief of the receiver as time $t\to\infty$ and number of sources $n \to \infty$. Let $\boldsymbol{\mu}_{t}(\theta) \in \mathbb{R}^{>0}, \theta \in \{0,1\}, \boldsymbol{\mu}_{t}(1)+\boldsymbol{\mu}_{t}(0) = 1$ denote the belief of receiver at time $t>0$. To study the evolution of belief, it is convenient to consider the $\log$ of belief ratio between the two states:  $\boldsymbol{\phi}_t = \log\left({\boldsymbol{\mu}_t(1)}/{\boldsymbol{\mu}_t(0)}\right)$. In this notation, if the true state of the world is $1$, then learning ($\boldsymbol{\mu}_t(1) \to 1$) is equivalent to  $\boldsymbol{\phi}_t \to\infty$. In such a regime, where learning happens asymptotically, we can write $\boldsymbol{\mu}_t(0) \asymp e^{-\boldsymbol{\phi}_t}$. In particular, if ${\boldsymbol{\phi}_t}$ grows linearly at rate $r$, then $\boldsymbol{\mu}_t(0)$ goes to zero asymptotically exponentially fast with the same (exponential) rate $r$. 

In the next section, we derive the log-belief ratio in the case where the receiver has perfect recall. Then, we explore deviation from this baseline with imperfect recall.

% \begin{align}
%     \boldsymbol{\phi}_t = \log\left({\boldsymbol{\mu}_t(1)}/{\boldsymbol{\mu}_t(0)}\right).
%     \label{eq:logbeliefratio}
% \end{align}

\subsection{The Bayesian Baseline}

Consider fixed $t>0$. The $\log$-belief ratio of the Bayesian receiver at $t$ is given by:
\begin{align}
    \boldsymbol{\phi}_t = \sum_{i=1}^{n}\mathds{1}(\mbox{i transmits at least once before  }t )\boldsymbol{\lambda}_i, \label{eq:bayesianlogratio}
\end{align} where $\boldsymbol{\lambda}_i = \mathbf{s_i}(\overline{\lambda}_i - \underline{\lambda}_i) + \underline{\lambda}_i$, and
\begin{align}
    \overline{\lambda}_i   = \log\left(\frac{\overline{p}_i}{\underline{p}_i}\right), \underline{\lambda}_i   = \log\left(\frac{1 - \overline{p}_i}{1 - \underline{p}_i}\right). %\label{eq:liklihooddef}
    \nonumber
\end{align}

Conditioned on the realizations of the private signals $\bar{\mathbf{s}} = (\mathbf{s}_1,\ldots,\mathbf{s}_n)$, the expected value of \eqref{eq:bayesianlogratio} is given by:
\begin{align}
    \mathbb{E}(\boldsymbol{\phi}_t|\overline{\mathbf{s}}) & = \sum_{i=1}^{n}\boldsymbol{\lambda}_i \mathbb{P}(i \mbox{ transmits at least once before }t ) \nonumber \\
    & = \sum_{i=1}^{n}\boldsymbol{\lambda}_i (1 - e^{-\alpha_i t}).  \label{eq:expectedbayesianlogratio}
\end{align} As $t \to\infty$, the Bayesian belief aggregates the $\log$-likelihood of the initial private signals perfectly: 
\begin{align}
    \mathbb{E}(\boldsymbol{\phi}^{\star}|\overline{\mathbf{s}}) := \lim_{t\to\infty} \mathbb{E}(\boldsymbol{\phi}_t|\overline{\mathbf{s}})  = \sum_{i=1}^{n}\boldsymbol{\lambda}_i.  \label{eq:expectedbayesianlogratiolimitTtoInfty}
\end{align}

\subsection{Imperfect Recall}
Consider now a non-Bayesian agent, $r>0$, that has an imperfect recall of the sources. After receiving a message $\mathbf{m}_{{\tau}} = (i,\mathbf{s}_i)$, she has to recall if the source $i$ has transmitted before according to \eqref{eq:proba_remember}. 

For fixed $t>0$, let $0< \boldsymbol{\tau_1} < \ldots < \boldsymbol{\tau_T} < t$ be the sequence of random time points at which messages are received from the sources. Let us also denote $\bar{\boldsymbol{\tau}} = (\boldsymbol{\tau_1}, \ldots , \boldsymbol{\tau_T})$. The $\log$-belief ratio of the non-Bayesian receiver at time $t$ can be written in terms of indicator functions, $\mathds{1}(\cdot)$, as follows:
\begin{align}
    \boldsymbol{\phi}_t & =  \sum_{j = 1}^{\mathbf{T}}\mathds{1}(\mbox{source }\mathbf{m}_{\boldsymbol{\tau}_j}^{(1)} \mbox{ not recalled at  }\boldsymbol{\tau_j} )\boldsymbol{\lambda}_{\mathbf{m}_{\boldsymbol{\tau}_j}^{(1)}} \nonumber \\
    & = \sum_{j = 1}^{\mathbf{T}}\sum_{i=1}^{n}\mathds{1}(i \mbox{ transmits at }\boldsymbol{\tau_j} )\mathds{1}(i \mbox{ not recalled at }\boldsymbol{\tau_j} )\boldsymbol{\lambda}_{i}. \nonumber
    %\label{eq:nonbayesianbeliefratio}
\end{align}

Conditioned on $\bar{\mathbf{s}}$ and $\bar{\boldsymbol{\tau}}$, the  expected $\log$-belief ratio of the non-Bayesian receiver at time $t$ is then given by:
\begin{multline}
 \mathbb{E}(\boldsymbol{\phi}_t|\bar{\mathbf{s}}, \bar{\boldsymbol{\tau}}) = \sum_{i=1}^{n}\dfrac{\alpha_i}{\bar{\alpha}_n}\boldsymbol{\lambda}_{i} \sum_{j = 1}^{\mathbf{T}}\mathbb{E} \bigg(1 - \\ \frac{|\tau\leq\boldsymbol{\tau}_j: \mathbf{m}_{{\tau}}^{(1)} = i|}{|\tau\leq\boldsymbol{\tau}_j: \mathbf{m}_{{\tau}}^{(1)} = i| + r|\tau\leq\boldsymbol{\tau}_j: \mathbf{m}_{{\tau}}^{(1)} \neq i|} \bigg).
 \nonumber
\end{multline}

Taking the limit $t \to \infty$, together with the ergodic limits of the Poisson point processes ($\mathbf{T}/(\bar{\alpha}_n t) \to 1$ and $|\tau\leq\boldsymbol{\tau}_j: \mathbf{m}_{{\tau}}^{(1)} = i|/(|\tau\leq\boldsymbol{\tau}_j: \mathbf{m}_{{\tau}}^{(1)} = i| + r|\tau\leq\boldsymbol{\tau}_j: \mathbf{m}_{{\tau}}^{(1)} \neq i|) \to \alpha_i/(\alpha_i + r(\bar{\alpha}_n - \alpha_i)$, almost surely as $t\to\infty$) yields:
    \begin{align}
    \lim_{t\to\infty}\mathbb{E}(\boldsymbol{\phi}_t|\bar{\mathbf{s}}, \bar{\boldsymbol{\tau}}) &=  \sum_{i=1}^{n} {\alpha_i} \boldsymbol{\lambda}_{i}  \left(
    1-\frac{\alpha_i}{ \alpha_i + r(\bar{\alpha}_n - \alpha_i)} \right) t.
    \nonumber
    %\label{eq:nonbayesianbeliefratio1}
\end{align} We can now drop the conditioning on the transmission times to get: 
\begin{align}
    \lim_{t\to\infty}\mathbb{E}(\boldsymbol{\phi}_t|\bar{\mathbf{s}}) &=  \sum_{i=1}^{n} {\alpha_i} \boldsymbol{\lambda}_{i} \left(1-\frac{\alpha_i}{ \alpha_i + r(\bar{\alpha}_n - \alpha_i)} \right)t.
    \nonumber
    %\label{eq:nonbayesianbeliefratio2}
\end{align} Hence, the belief of the non-Bayesian agent concentrates exponentially fast as $t \to\infty$ and the rate of convergence is given by:   
\begin{align}
    \bar{{\boldsymbol{\phi}}}_n & :=  \lim_{t\to\infty}\frac{1}{t}\mathbb{E}(\boldsymbol{\phi}_t|\bar{\mathbf{s}}) \nonumber \\
    & = \sum_{i=1}^{n} {\alpha_i} \boldsymbol{\lambda}_{i} \left( 1-\frac{\alpha_i}{ \alpha_i + r(\bar{\alpha}_n - \alpha_i)} \right). 
    \label{eq:nonbayesianbeliefratio3}
\end{align}
As $t\to\infty$, the non-Bayesian belief concentrates on one if $\bar{{\boldsymbol{\phi}}}_n > 0 $ and it concentrates on zero if $\bar{{\boldsymbol{\phi}}}_n < 0$. We list three consequences of the preceding limit behavior as follows. First, Non-Bayesian receiver adopts more extreme beliefs as $t \to \infty$. This is in spite of her limited information (there are only $n$ signals available to her), and in contrast to the Bayesian belief in \eqref{eq:expectedbayesianlogratiolimitTtoInfty}, which aggregates the $n$ signals perfectly. Second, The non-Bayesian belief may concentrate on the wrong state. In fact, if the two states are equally likely, then the ex ante probability of non-Bayesian belief concentrating on a wrong state is given by $1/2\mathbb{P}\left(\bar{{\boldsymbol{\phi}}}_n < 0 \,\middle\vert\, \theta = 1\right) + 1/2\mathbb{P}\left(\bar{{\boldsymbol{\phi}}}_n > 0 \,\middle\vert\, \theta = 0\right)$. Third, the non-Bayesian belief is disproportionately influenced by different sources according to their transmission rates. This is in contrast to the Bayesian belief in \eqref{eq:expectedbayesianlogratiolimitTtoInfty} that weighs all signal log-likelihood ratios equally. In the absence of coordination or knowledge of signals from other sources, source $i$ shapes the opinion of the receiver both directly (through $\alpha_i/(1-\alpha_i/(\alpha_i + r(\bar{\alpha}_n - \alpha_i))$) and indirectly (through $r\bar{\alpha}_n$ which affects all other terms). In the appendix, we analyze the implication for pricing to regulate bandwidth and learning in limited environments when $r=1$ (all signals interfere as much as signals from source $i$ when recalling source $i$) and total bandwidth (the denominator) is fixed at $B$ so that the influence of $i$ can be written as $\alpha_i(1-\alpha_i/B)$.

\section{Experiment}

The previous section quantified the influence of the ith source in shaping the opinion of the receiver as a function a parameter $r$ which can be interpreted as the strength of interference from other senders. How large is $r$, and therefore the influence of the $i$th source in practice? To estimate our model, we run an experiment where participants observe a feed aggregating comments about an issue following which they need to estimate the number of people who hold one opinion or the other.

\subsection{Experimental Setup}
Our setup is based on the famous picture of ``The Dress'' that generated significant disagreements when it first came out in February 2015 \citep{lafer2015striking}. Different people see different colors: some people see the dress as black and blue; others  see it as white and gold. In fact, a poll was run on the website Buzzfeed asking users what color is this dress\footnote{The original poll can be found at \url{https://www.buzzfeed.com/catesish/help-am-i-going-insane-its-definitely-blue\#subbuzz-quiz-poll-results-5103755}}. We scraped all comments from this website and showed respondents a random sample.

\begin{figure*}[h]
 \centering
	\includegraphics[width=13cm]{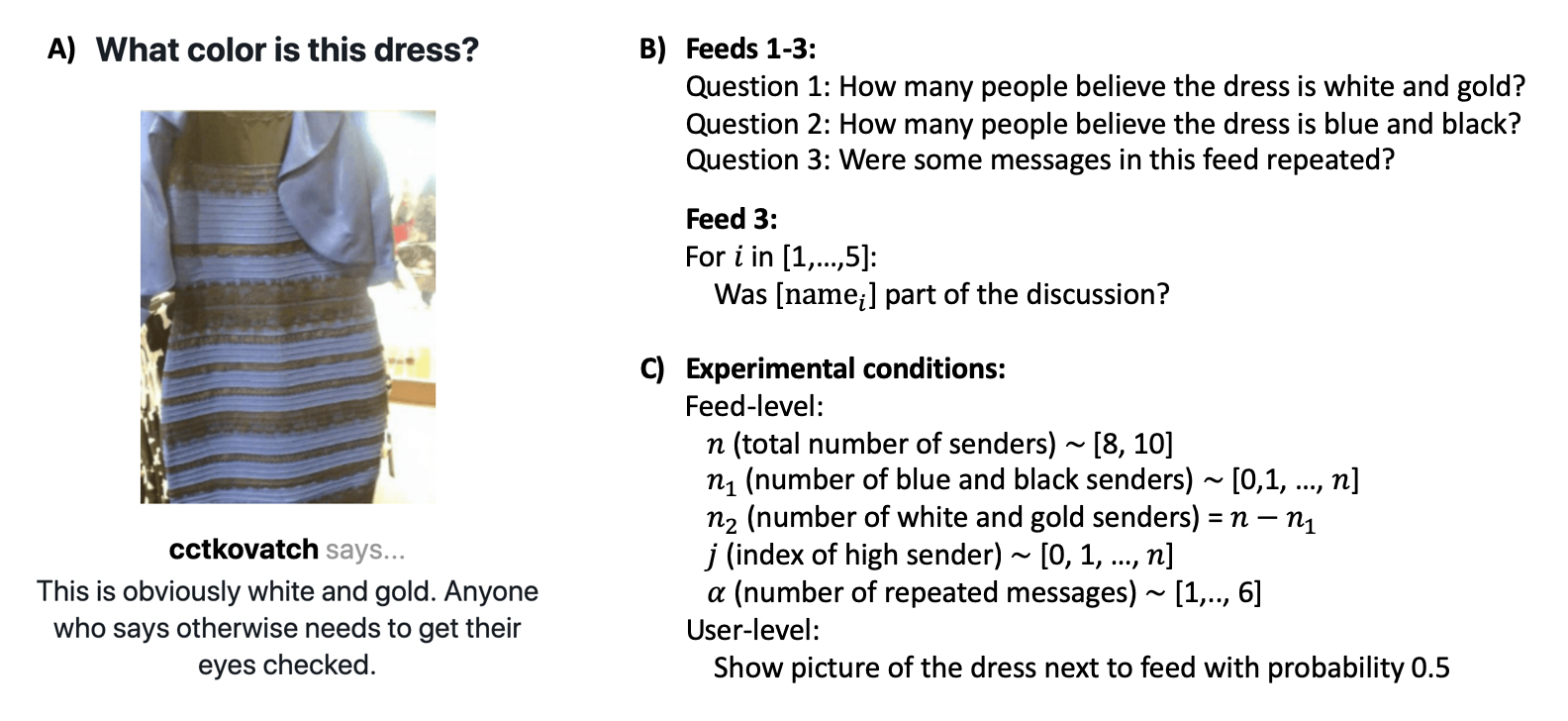}
	\caption{A) Picture of the ``The Dress" as was shown in the experiment. Under the picture is a username along with the message. During the experiment, the username appears first for two seconds, after which the message is shown for an additional five seconds. B) Questions asked after each feed (note that the order of questions one and two is randomized). C) Experimental Conditions. The experiment was coded using the Empirica framework \citep{almaatouq2021empirica}.}
	\label{fig:thedress}
	\centering
\end{figure*}

Figure  \ref{fig:thedress} shows how this was translated in our experiment. A picture of the dress is shown with comments appearing at the bottom. To make sure respondents  pay attention to the username, we make it appear first for two seconds, following which we add the comments (as shown in the picture) for 5 seconds. We only show one message at a time because some of them will be repeated, and we want to avoid the possibility of showing the same message multiple times on the same page.

\subsection{Experimental Conditions}
Participants see a total of three different feeds. In the first feed, participants do not know what they will be asked. They are simply told that they will be asked a series of questions about what they have just seen. After they answer the question at the end of the first feed, the same setup is repeated in a second and third feed. This design choice has two main advantages. First, it serves as training: the main analysis will be performed on the second and third feed, so the first one allows respondents to know what to expect. Second, it allows us to compare how people respond when they know the question relative to when they don't. In fact one premise of this work  is that people are not constantly updating their counts in real time [in a similar vein to \citet{enke2020associative}]. Depending on the environment in which they are located --- e.g., if they are reading product reviews --- they may in fact be doing so. However, in other environments such as social media feeds, people do not update their counts unless they are expected to make a decisions (upon which they need to recollect all the information that has been received  up until that point).

The next set of experimental variation occurs at the feed level, leading to within (in addition to between) subject randomization. Let us denote by $n_1$ and $n_0$ the number of senders who send a message corresponding to `blue and black' and `white and gold' respectively. $n_1$ is drawn uniformly from $[0, n]$ where $n$ is the total number of senders randomly chosen to be $8$ or $10$. $n_0$ is simply $n-n_1$. The high sender is randomly chosen across all senders and the number of repeated messages, denoted by $\alpha$, is drawn from $[1,6]$. Finally, next to each feed is a picture of the dress which we decide to show (or not) with equal probability.  This randomization occurs at the user level. 

\subsection{Questions and Incentive Compatibility}
The first question we ask is: ``what color do you see?". We ask this question at the beginning of the experiment (after the instructions, before they observe the first feed) to see how subsequent answers vary according to whether the participant agrees (or disagrees) with the high sender. 

Then, we ask the same three questions following each feed. First, we ask participants to estimate the number of people who believe the dress is `black and blue'. Second, we ask them to estimate the number of people who believe the dress is `white and gold'. (Note that whether we ask `black and blue' or `white and gold' first is randomized.) Third, we ask participants if some messages were repeated. Finally, at the end of the last feed only, we ask participants if they remember seeing each of five names (the high sender, one random low sender who sent `black and blue', one random user who sent `white and gold' and two senders that were not shown). Each name appears on a separate page and the order of the names is randomized. We ask this at the end of the last feed only (not the first or second) to ensure that respondents focus on counting (as they would if they were trying to aggregate opinions) not remembering names.

Participants are paid according to performance. They start from a base amount that depends on the total number of messages they are assigned to observe. An additional (constant) amount is added/deduced for each answer depending on whether it is correct/incorrect.

\subsection{Ethics and Protection of Human Subjects}
This protocol was determined to be exempt by the MIT Committee on the Use of Humans as Experimental Subjects. The decision to participate in this experiment was entirely voluntary. There were no known or anticipated risks to participating in this experiment. There was no way for us to identify participants. The only information we collected, in addition to participants' responses, is the timestamps of their interactions with our site. These points were listed at the beginning of the experiment, along with the statement that the results of this research may be presented at scientific meetings or published in scientific journals. Before starting, participants had to click on the ``AGREE" button, indicating that they are at least 18 years of age, and agree to participate voluntarily.

\subsection{Setup Strength}
This experimental setup has two main advantages. By showing respondents comments that were scraped from the internet, this setup aims to provide respondents with a realistic experience. In addition, the nature of the perceptual differences implies that repeated messages contain no additional informational content. In fact, one of the biggest issues with over-counting is that repetition may be interpreted by the respondent as evidence that the sender has learned something new about the world or that his opinion has changed. For example, one's opinion of a painting may change over repeated exposures. Here, the picture is such that the color people see remains fixed throughout. As a result, seeing twice the message ``I see the color white and gold'' cannot be rationally interpreted as more than a single piece of evidence.

\subsection{Estimation}

Let $x^{j,k}$ correspond to the number of messages sent by the high sender in feed $j$ that participant $k$ is exposed to. Let us denote by $n_1^{j,k}$ the set of users who agree with the high sender (and by $n_0^{j,k}$ those who disagree with him). Finally, let $Y_i^{j,k}$ be  the estimated size of $n_i^{j,k}$. Therefore,
\begin{align}
Y_0^{j,k} = |n_0^{j,k}| + u_0^{j,k},
\nonumber
\end{align}
where $u_0^{j,k}$ captures all the unobserved factors affecting $k$'s performance estimating the number of senders who disagree with the high sender in feed $j$.

$Y_1^{j,k}$ depends on the fraction of repeated messages $\alpha^{j,k}$ that will be over-counted. This fraction is equal to $1-p_r$ where $p_r$ is the probability of remembering the high sender as defined in \eqref{eq:proba_remember}:
$$
p_r=\dfrac{\alpha^{j,k}}{\alpha^{j,k} + r(\bar{\alpha}^{j,k} - \alpha^{j,k})},
$$
where $\bar{\alpha}^{j,k}$ corresponds to the total number of messages sent and $r$ is the strength of interference. 
Therefore,
\begin{align}
Y_1^{j,k} = |n_1^{j,k}| + \alpha^{j,k}(1-p_r) + u_1^{j,k}.
\nonumber
\end{align}
Notice that this is exactly equal to the  influence of the high sender on the belief of the receiver, cf. \eqref{eq:nonbayesianbeliefratio3}.\footnote{Note that, in this case, it is also equal to the direct influence (because the indirect influence is zero, as there is no repetition from other senders).} Therefore, our setup allows us to estimate the parameter of interest while abstracting away from the Bayesian update.

We further assume that $u_i^{j,k}$ can be separated in two terms, namely $u_i^{j,k}=\eta^{j,k}+\epsilon_i^{j,k}$ where $\epsilon_i^{j,k}$ captures all the unobserved factors affecting $k$'s performance estimating the number of $i$ senders in feed $j$ (such as the inability to read a specific message) and $\eta^{j,k}$ captures the unobserved factors common to $i=0$ and $i=1$ (such as lack of attention or tiredness). The distinction between these two types of unobserved factors justifies taking the difference between $Y_1^{j,k}$ and $Y_0^{j,k}$ to remove the $\eta^{j,k}$ term.
We therefore assume 
\begin{align}
Y^{j,k} = \alpha^{j,k}(1-p_r) + \epsilon^{j,k} 
\label{eq:main_estimation}
\end{align}
where $Y^{j,k}=(Y_1^{j,k} - Y_0^{j,k})  -  (|n_1^{j,k}|-|n_0^{j,k}|) $, and $\epsilon^{j,k} = \epsilon_1^{j,k} - \epsilon_0^{j,k}$. The parameter $r$ will be estimated with MCMC using the Stan framework \citep{stan} assuming that $\epsilon^{j,k} \sim N(0, \sigma_{\epsilon})$.

\section{Results}

We estimate equation \ref{eq:main_estimation} using data collected from 1444 online workers, 416 of which were removed according to the following pre-registered criteria. First we performed a simple attention check, removing all respondents who failed to answer ``what is 2+2?'' and ``what is the color or Napoleon's white horse?''. Second, we removed from the analysis all the answers that differed by more than 5 from the true answer in the second and third feeds (i.e. when they know the question). This suggests that respondents may be answering randomly. We selected US participants from Amazon Mechanical Turk with a HIT approval rate higher than 95\% and a number of HITs approved higher than 500.

\subsection{Memory Checks}
\begin{figure}[tb]
    \centering
	\includegraphics[width=6.5cm]{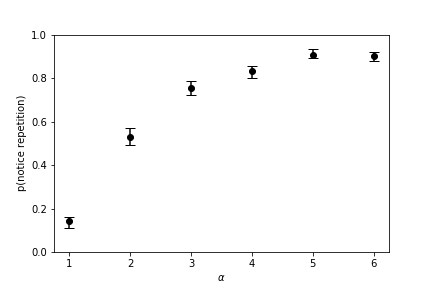}
	\caption{Probability of noticing repetition coming from answers to the question ``Were some messages in this feed repeated?" in the second and third feeds as a function of $\alpha$, the number of repeated messages. The 95\% confidence intervals were computed using a bootstrap clustered at the user level. Each dot corresponds to the median}
	\label{fig:p_notice}
\end{figure}

Before running the analysis, we verify that participants do indeed recognize the presence of repeated messages. Figure~\ref{fig:p_notice} shows how the probability of `yes' answers to the question ``Were some messages in this feed repeated?'' increases with $\alpha$ in the second and third feeds. Consistent with our expectation, this probability increases rapidly until it reaches a maximum of 0.9 when $\alpha=5$. This question was also asked in the first feed so participants were incentivized to perform well.

\begin{figure}[tb]
    \centering
	\includegraphics[width=6.5cm]{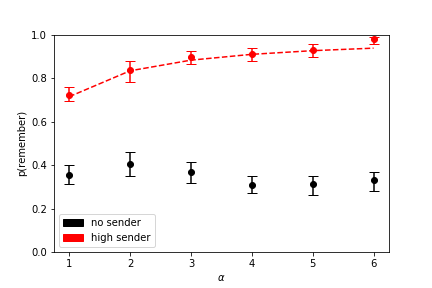}
	\caption{
	Probability of remember a ``no sender'' (i.e., incorrectly remembering a name that was not shown on the feed, in black) and the high sender (who repeats himself $\alpha$ times, in red) as a function of $\alpha$, the number of repeated messages. The 95\% confidence intervals were computed using a bootstrap clustered at the user level.}
	\label{fig:p_remember}
\end{figure}

We next turn to the memory question asked at the end of the third feed. Figure \ref{fig:p_remember} shows the probability of remembering the high sender (who repeats himself $\alpha$ times, in red) and the probability of remembering someone whose name was not shown in the feed (in black) as a function of $\alpha$. As the graph suggests, the probability of remembering the high sender as a function of repetition is consistent with the probability model introduced. In fact, estimation of equation \eqref{eq:proba_remember} leads to $\hat{r}=0.05$ (95\% CI: [0.05, 0.06]) and prediction of the model is plotted as the red dotted line. However, note that these questions were not incentivized, so we expect this estimate of $r$ to be different from the true value. Seeing that the probability of remembering a name absent from the feed is positive and statistically different from zero, we expect the bias to be positive, i.e. probabilities would be lower if this task was incentivized and therefore $r$ would be higher.

\subsection{Empirical Analysis}

\begin{figure}[tb]
 \centering
	\includegraphics[width=6.5cm]{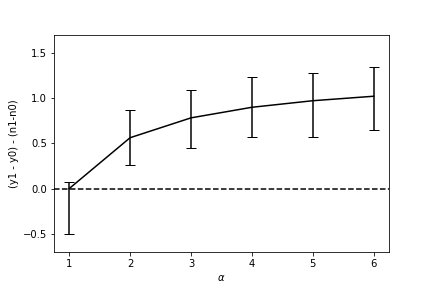}
	\caption{Estimated deviation from the truth $(y_1 - y_0) - (n_1-n_0)$ as a function of $\alpha$, the number of repeated messages from the high sender. The total number of senders is 8. 95\% confidence intervals are computed using a bootstrap clustered at the user level. The line passing through the intervals corresponds to the prediction of the model.}
	\label{fig:main2}
	\centering
\end{figure}

We start our empirical analysis by estimating equations \ref{eq:main_estimation} using data from the second and third feed. Figure \ref{fig:main2} shows how the main outcome of interest varies with $\alpha$ when the total number of senders is 8. The main outcome of interest $(y_1 - y_0) - (n_1-n_0)$ which represents deviation from the truth (the estimated difference minus the true difference) is centered at 0 when $\alpha=1$, as expected. However, as repetition increases above one, respondents increasingly deviate from the $y=0$ line (what the answer would be if they had perfect memory).

Estimation of equation \ref{eq:main_estimation} leads to $\hat{r}=0.16$ (95\% CI: [0.12, 0.19]). As expected, this is higher than the estimated $r$ when the task is not incentivized. We then investigated whether the effect depends on whether the participant sees the same color as the high sender or not, i.e., we separated $r=r_0+\mathds{1}(\text{same color})r_1$. Resulting estimates are $\hat{r}_0=0.12$ (95\% CI: [0.08, 0.16]) and $\hat{r}_1=0.09$  (95\% CI: [0.02, 0.17]). So perceiving the same color increases the strength of interference from other senders ($t=81.8, p<10^{-5}$). One interpretation for this result is that disagreement with the high sender leads to surprise, which makes participants more attentive.
We replicated this analysis by including data from the first feed, i.e. when respondents know that they need to be attentive but they do not know what they will be asked. Results show that not knowing the question increases interference by 0.17 (95\% CI: [0.05, 0.3]).

%\begin{figure}[h]
%	\includegraphics[width=8cm]{same_color}
%	\caption{Estimated strength of interference when the participant sees the same color as the high sender (right) or when the participant sees the opposite color (left). Parameters were estimated with MCMC using the Stan framework. 95\% confidence intervals are shown.}
%	\label{fig:same_color}
%	\centering
%\end{figure}

We also looked at whether this effect varies by education, but in this case $r_1$ was not statistically different from zero, which suggests that, conditional on being selected to perform the task, education has no effect.

%\begin{figure}[h]
%	\includegraphics[width=8cm]{question}
%	\caption{Estimated strength of interference when the participant does not know the question (left, which corresponds to feed 1) or when the participant knows the question (right, which corresponds to feeds 2 and 3). 95\% confidence intervals are shown}
%	\label{fig:question}
%	\centering
%\end{figure}

\subsection{Related Work}
Memory constraints have been  looked at in the context of social learning \cite[ch. 5]{Chamley2004}. In recent results, \citet{wilson2014bounded} considers the model of a decision maker who chooses between two actions with pay-offs that depend on the true state of the world. Furthermore, the decision maker must always summarize her information into one of finitely many states, leading to optimal decision rules that specify the transfers between states. \citet{Kocer:2010:ELB:2049103} extends this set up to more complex, dynamic learning environments including partially observed Markov decision processes and multi-armed bandits.

The problem of learning with finite memory in the context of hypothesis testing was originally formulated by \citet{cover1968note,cover1969hypothesis} under memory constraints for storing the test statistics. Accordingly, while sufficient statistics are very useful computational tools their utility for memory reduction is not clear. Subsequent results provide sophisticated algorithms using automata to perform the task of hypothesis testing using test statistics that take only finitely many values and guarantee an asymptotically vanishing error probability \cite{Kontorovich2012,hellman1970learning,Cover197649,cover1970two}.

More recently,  \citet{drakopoulos2013learning} have considered this problem in a setting where agents each receive an independent private signal and make decisions sequentially. Memory in this context refers to the number of immediate predecessors whose decisions are observable by any given agent at the time of making her decision. Accordingly, while the almost sure convergence of the sequence of individual decisions to the correct state is not possible in this finite memory setting, the authors construct decision rules that achieve convergence and learning in probability. They next go on to consider the behavior of rational (pay-off maximizing) agents in this context and show that in no equilibrium of the associated Bayesian game learning can occur. Our work is complementary to these studies. We are interested in bounded memory effects in so far as the ability of the receiver to recall information sources (or identity of the senders) is concerned. We model the recall process based on empirically validated assumptions of cued recall, using the concept of interference.

Related to recall of sources, \citet{acemoglu2014dynamics} analyze social learning among agents who directly communicate their entire information sets (represented as pairs of private signals and their sources). Since each piece of information is tagged there is no confounding and Bayesian updating is simple. They show that the presence of information hubs and social connectors facilitates efficient learning outcomes in the communicative model. This is in contrast to some of the insights from the observational models \cite{BalaGoyal,GolubWisdomCrowd}, where learning from actions of neighbors is impeded by the presence of highly connected, influential agents. In this paper, we are interested in the agents that have imperfect recall of the sources. On the other hand, our learning environment is much simpler. The sources repeat their messages; therefore,  two messages are confounded only if they are repeated messages from the same sources.

\section{Discussion}

This paper investigated the effect of repetition on learning through the lens of limited memory. To study limited recall, we proposed a simple model that is both analytically tractable and accounts for two empirically validated facts of human memory \citep{kahana2012foundations}. Repetition increases the likelihood of recall. However, it is constrained by what we call interference: messages from other senders decrease the probability of remembering any single message. We used this model to derive an analytic formulation allowing us to quantify the influence of each sender, as a function of a parameter $r$ that can be interpreted as the strength of interference. We estimated this parameter in an online experiment where participants need to count signals from different senders. Our results provide evidence that $r$ is positive and is decreased if participants disagree with the source of repetition, which we attribute to increased attention. Further, $r$ is decreased if participants know beforehand the question they are incentivized to answer correctly.

This experiment leveraged a unique feature of a picture, ``The Dress", that naturally separates people into those who see the color as blue and black and those who see it as white and gold. Importantly, since the perceived color does not change over repeated exposures, the effect of repetition can be estimated by abstracting away from all other factors. In other settings, e.g., political questions where people may have an existing belief, the effect of repetition is reinforced by two facts. First, people are not incentivized to be attentive to repetition (as they are in our experiment). Second, it may induce opinion changes [e.g., though the ``illusory truth effect" \citep{hasher1977frequency, fazio2019repetition, hassan2021effects}]. Our results can therefore be interpreted as a lower bound for the total effect of repetition in typical news feed designs.

Our results have implications for opinions and decisions that depend on the opinion of others \citep{degroot1974reaching, demarzo2003persuasion}. Through repetition, influential agents may induce their network neighbors to make the wrong choice --- by leading them to overestimate the number of people who share their opinion. This suggests a natural policy where platforms keep track of repetition and notify users when they are exposed to repeated information.

This policy may not be as effective when people are not constantly updating their beliefs in real-time [in a similar vein to \citet{enke2020associative}]. In many environments, people may not update aggregate counts unless they are expected to make a decision (upon which they need to recollect all the information that has been received up until that point). In such cases, people may learn to rationally avoid notifications. Consequently, a more effective strategy may be to hide repeated information to reduce extreme opinion formation in online platforms. An additional difficulty stems from the fact that, contrary to our experiment, repeated signals are rarely exactly identical (although they may have the same meaning), which makes the problem of detecting repetition, and therefore overcounting, harder for humans and machines. This provides a unique opportunity for platforms to construct similarity metrics between signals to help humans discard repeated information.

While this paper estimated the role of repetition from a single high sender, we see many promising opportunities for further study of competition effects from other high senders. An implication of model \ref{eq:proba_remember} is that repetition from a sender increases interference (through the denominator), which means that repetition from other senders is less likely to be noticed. Another extension would be to analyze how the effect of repetition varies by sender. Different senders may vary according their popularity or how close they are to the low sender. One could incentivize some high senders to increase the fraction of repeated messages and explore how influence varies based on characteristics of the sender, receiver or their relationship. Finally, we could implement a policy that hides a random fraction of repeated messages in a real social network to quantify how influence varies with repetition.

\section{Acknowledgments}
M.A.R. acknowledges support from Pitt Momentum Funds and a Pitt Cyber Accelerator grant. D.E. acknowledges support from an Amazon Research Award and a MIT Sloan Junior Faculty Research grant. D.E. previously, while contributing to this research, had a significant financial interest in Facebook and is a consultant to Twitter. D.E. has received funding for other research from Facebook, and Facebook has funded a conference he organizes.

\appendix

\section{Appendix}
\subsection{Pricing to Regulate Bandwidth Usage}

We consider a scenario where each source is interested in maximizing their influence while they pay a cost for increased transmission rates. To simplify the analysis, let us assume that $r=1$ such that the influence of the $i$-th source in shaping the opinion of the receiver (in the absence of coordination or knowledge of signals from other sources) is $\alpha_i/(1-\alpha_i/\bar{\alpha}_n)$

Accordingly, sender $i$ chooses the rate $\alpha_i$ to maximize:
\begin{align}
f(\alpha_i|p(\mathord{\cdot}), {\alpha}_{-i}) &= {\alpha_i} \left( \frac{\bar{\alpha}_n - {\alpha_i}}{ \bar{\alpha}_n} \right) - p(\alpha_i) \nonumber
\\ &=  {\alpha_i} \left( \frac{\bar{\alpha}_{-i}}{ \bar{\alpha}_{-i} + {\alpha}_{i}} \right) - p(\alpha_i), \label{eq:utility}
\end{align} where $p(\alpha_i)$ is the price that the platform changes to sender $i$ for transmission at rate $\alpha_i$, and  ${\alpha}_{-i} = (\alpha_1,\ldots, \alpha_{i-1},\alpha_{i+1},\ldots,\alpha_n)$ is the rate profile chosen by all the senders other than $i$, and $\bar{\alpha}_{-i} = \sum_{j\neq i} \alpha_j$ is the sum of the transmission rates of all senders other than $i$. 

To proceed, we further assume that $\bar{\alpha}_n = B$ is fixed with $n$. This is motivated by practical concerns in platforms: To enhance the user experience, the platform would like to fix the total rate at which a user receives messages --- one way to keep the total receiving rates fixed is by dropping messages sufficiently fast (homogeneously across sources), cf. Subsection \ref{sec:limittedbandwidth}. Here, we consider a situation where senders choose their rates optimally in response to the price function charged by the platform. With fixed $\bar{\alpha}_n = B$, the sender's utility is given by:  

\begin{align}
f(\alpha_i|p(\mathord{\cdot}), B) = {\alpha_i} \left( \frac{B - {\alpha_i}}{B} \right) - p(\alpha_i),\label{BfixedUtility}
\end{align}

The price function that induces ${\alpha}_i = \alpha^{\star} = B/n$ for all $i$ satisfies: $1 - 2/n - p'(B/n) = 0$. For example, with a linear cost structure $p(\alpha) = c_1\alpha$ we get: $c_1 = 1-2/n$, and with a quadratic cost structure $p(\alpha) = c_2\alpha^{2}$ we obtain: $c_2 = (n-2)/(2B)$. Using such pricing strategies the platform can regulate the bandwidth usage to ensure equatable distribution of rates. In fact, the price structure can depend on different observable source features (in addition to rate) and may be applied to guarantee other platform-desirable outcomes.

\subsection{Learning in Limited Bandwidth Environments}\label{sec:limittedbandwidth}

We begin by considering the Bayesian case. Averaging out the private signals in \eqref{eq:expectedbayesianlogratio} gives:
\begin{align}
    \mathbb{E}(\boldsymbol{\phi}_t) & = \sum_{i=1}^{n}D_{KL}(\overline{p}_i||\underline{p}_i)(1 - e^{-\alpha_i t}),  \label{eq:expectedbayesianlogratiosignalsaveragedout}
\end{align} where  
\begin{multline}
    D_{KL}(\overline{p}_i||\underline{p}_i) = \mathbb{E}(\boldsymbol{\lambda}_i) = \\ \overline{p}_i \log\left(\frac{\overline{p}_i}{\underline{p}_i}\right) + (1 - \overline{p}_i) \log\left(\frac{1 - \overline{p}_i}{1 - \underline{p}_i}\right),
\end{multline} is the binary relative entropy between the signal distributions in each of the two states.

The limiting $\log$-belief ratio as $n\to\infty$ is given by:

\begin{align}
    \mathbb{E}(\boldsymbol{\phi}_t)  = \sum_{i=1}^{\infty}D_{KL}(\overline{p}_i||\underline{p}_i)(1 - e^{-\alpha_i t}).  \label{eq:expectedbayesianlogratioLIMITnTOinfty}
\end{align} By the law of large numbers, $\boldsymbol{\phi}_t  \asymp \sum_{i=1}^{n}D_{KL}(\overline{p}_i||\underline{p}_i)(1 - e^{-\alpha_i t})$, almost surely, as $n \to\infty$. In particular, for any fixed $t>0$, Bayesian learning in this environment is asymptotically exponentially fast in $n$ and the exponential rate of learning (with increasing $n$) is given by: 
\begin{align}
    {\phi}^\star_t := \lim_{n\to\infty}\frac{1}{n} \sum_{i=1}^{n}D_{KL}(\overline{p}_i||\underline{p}_i)(1 - e^{-\alpha_i t}) > 0.  \label{eq:expectedbayesianlogratioLIMnnINFTY}
\end{align} We can interpret the positivity condition in the right hand side of \eqref{eq:expectedbayesianlogratioLIMnnINFTY} as the learning criterion: \emph{For any fixed $t>0$ if ${\phi}^\star_t>0$, then the Bayesian agent aggregates the information from infinitely many senders and learns the true state of the world}. Taking further the limit $t \to \infty$, gives the following Bayesian Benchmark for the learning rate with the increasing population of the senders ($n$):
 \begin{align}
    {\phi}^\star := \lim_{t\to\infty} {\phi}^\star_t = \lim_{n\to\infty}\frac{1}{n} \sum_{i=1}^{n}D_{KL}(\overline{p}_i||\underline{p}_i).  \label{eq:expectedbayesianlogratioLIMttINFTY}
\end{align}

If we are in a regime with limited bandwidth (it is often the case that platforms are limited in the number of messages that they can show to the users), then the transmission rates should decrease with the increasing number of users. This decrease can be imposed exogenously by the platform. Here, we consider a setup where the platform drops the messages with probability $1 - p_n$ independently at random. The parameter $p_n$ is set such that the total rates at which the receiver (the platform user) receives messages is fixed and equal to $B_n$; hence, $p_n = B_n/\bar{\alpha}_n$. For the limited bandwidth environment to be interesting (non-trivial), we assume that $B_n < \bar{\alpha}_n$.  

To proceed, let $\bar{\alpha} = \lim_{n\to\infty}{\bar{\alpha}_n}/{n}$ be a population parameter that is the average transmission rate of the senders and is assumed fixed. Then $p_n = B_n/n \bar{\alpha}$ and the rate at which the receiver receives messages from each source $i$ is scaled down by $p_n$: $\alpha_{i}^{(n)} = p_n \alpha_i$; hence \eqref{eq:expectedbayesianlogratiosignalsaveragedout} becomes:
\begin{align}
    \mathbb{E}(\boldsymbol{\phi}_t) & = \sum_{i=1}^{n}D_{KL}(\overline{p}_i||\underline{p}_i)(1 - e^{-\alpha^{(n)}_i t}),  \label{eq:expectedbayesianlogratiosignalsaveragedoutNdependent}
\end{align} We can still ensure that the learning criteria in \eqref{eq:expectedbayesianlogratioLIMnnINFTY} is satisfied by requiring that $t_n = t \to \infty$, fast enough with increasing $n$. In fact, it suffices to have:  $\lim_{t_n \to\infty}(1 - e^{-\alpha^{(n)}_i t_n})  = \Omega(1) $ or $\lim_{t_n \to \infty}e^{- p_n \alpha_i t_n} = o(1)$. Hence, a sufficient condition for learning in bounded bandwidth settings is: $(B_n\alpha_i/n \bar{\alpha})  t_n  = \omega(1)$ or $t_n = \omega(n/B_n)$. Rather than at any $t>0$, the learning criterion is now satisfied only for large enough $t_n$. Hence, \emph{the effect of bounding bandwidth in large populations is to slow down the Bayesian learning process until the agent receives enough signals from the large population}. 

We can now consider the effect of bounding the bandwidth on learning with imperfect source recall. Again we begin by averaging out the initial signals. The expectation of the asymptotic belief in \eqref{eq:nonbayesianbeliefratio3} is as follows:

\begin{align}
    \bar{{\phi}}_n := \mathbb{E}(\bar{{\boldsymbol{\phi}}}_n) =  \sum_{i=1}^{n} {\alpha_i} D_{KL}(\overline{p}_i||\underline{p}_i) \left( \frac{\bar{\alpha}_n - {\alpha_i}}{ \bar{\alpha}_n} \right). 
    \label{eq:nonbayesianbeliefratio4}
\end{align}The learning criterion for the non-Bayesian receiver as $n$ goes to $\infty$ can be expressed as follows:$ \lim_{n\to\infty}  \sum_{i=1}^{n} {\alpha_i} D_{KL}(\overline{p}_i||\underline{p}_i)  = +\infty$. Here we have assumed that sender transmission rates are fixed with $n$ so that $\lim_{n\to\infty}\left( 1 - {\alpha_i}/{ \bar{\alpha}_n} \right) = 1$ for all $i$. Similar to \eqref{eq:expectedbayesianlogratioLIMttINFTY} we can introduce the following exponential rate for non-Bayesian learning with the increasing population of senders ($n$):
 \begin{align}
    \hat{{\phi}} := \lim_{n\to\infty}\frac{1}{n} \sum_{i=1}^{n}{\alpha_i} D_{KL}(\overline{p}_i||\underline{p}_i).  \label{eq:expectedbayesianlogratioLIMttINFTYnonBayesian}
\end{align}

Next consider the case where the platform drops messages independently at random with probability $1 - p_n = 1 - (B_n/n \bar{\alpha})$ to maintain a Bandwidth $B_n$. The learning criterion for the non-Bayesian receiver becomes:
\begin{align}
    \lim_{n\to\infty}  \frac{B_n}{n\bar{\alpha}}\sum_{i=1}^{n} {\alpha_i} D_{KL}(\overline{p}_i||\underline{p}_i) = \lim_{n\to\infty}  \frac{\hat{{\phi}}}{\bar{\alpha}} B_n  =  +\infty. 
    \label{eq:nonbayesianlearningcriteriap_nNonBayesian}
\end{align} While the Bayesian receiver learns after a long enough time in any limited bandwidth environment (as long as $t_n = \omega(n/B_n)$), the non-Bayesian receiver learns only if the bandwidth increases with the increasing population size: $\lim_{n\to\infty} B_n = \infty$. Moreover, for the non-Bayesian learning to occur asymptotically (at least) exponentially fast with the sender population size, we need: $\lim_{n\to\infty}B_n/n > 0$ or $B_n = \Omega(n)$. In particular, if  $0<\bar{B} = \lim_{n\to\infty}B_n/n < \bar{\alpha}$, is fixed, then the platform keeps a constant fraction of messages at random with probability $p = \bar{B} /\bar{\alpha}$ and the asymptotic exponential rate of non-Bayesian learning with the increasing population size is equal to: ${\bar{B}\hat{{\phi}}}/{\bar{\alpha}}$. 

Interference puts a major limitation on the ability of the non-Bayesian learner in limited bandwidth environments: \emph{Unlike the Bayesian receiver, a non-Bayesian receiver cannot learn from increasing population sizes, unless the bandwidth increases with the population size}.

\bibliography{ref.bib}

\end{document}